\documentclass[preprint,aps,prl,superscriptaddress]{revtex4-1}

\usepackage{amsmath}    
\usepackage{breqn}      
\usepackage{graphicx}   
\usepackage{verbatim}   
\usepackage{color}      
\usepackage{subfigure}  
\usepackage{hyperref}   
\newcommand{\ket}[1]{\ensuremath{\lvert #1 \rangle}}
\newcommand{\bra}[1]{\ensuremath{\langle #1 \rvert}}

\newcommand{\ald}{\ensuremath{\hat{a_l}^{\dagger}}}
\newcommand{\al}{\ensuremath{\hat{a_l}}}
\newcommand{\bjd}{\ensuremath{\hat{b_j}^{\dagger}}}
\newcommand{\bj}{\ensuremath{\hat{b_j}}}

\newcommand{\olm}{\ensuremath{\omega_l}}
\newcommand{\oj}{\ensuremath{\omega_j}}
\newcommand{\oa}{\ensuremath{\omega_a}}
\newcommand{\ob}{\ensuremath{\omega_b}}
\newcommand{\bt}[1]{\ensuremath{\hat{B_j}\left(t_{#1}\right)}}

\newcommand{\ave}[1]{\ensuremath{\langle #1 \rangle}}

\newcommand{\second}{\ensuremath{g^{(2)}}}

\begin{document}

\title{Determining Phonon Coherence Using Photon Sideband Detection}
\author{Ding Ding}
\affiliation{Department of Mechanical Engineering$,$ University of Colorado Boulder$,$ Boulder$,$ CO 80309$,$ USA }
\affiliation{Singapore Institute of Manufacturing Technology$,$ 2 Fusionopolis Way$,$ Singapore 138634}
\author{Xiaobo Yin}
\affiliation{Department of Mechanical Engineering$,$ University of Colorado Boulder$,$ Boulder$,$ CO 80309$,$ USA }
\author{Baowen Li}
\email{Baowen.Li@colorado.edu}
\affiliation{Department of Mechanical Engineering$,$ University of Colorado Boulder$,$ Boulder$,$ CO 80309$,$ USA }


\date{\today}

\begin{abstract}
Generating and detection coherent high-frequency heat-carrying phonons has been a great topic of interest in recent years. While there have been successful attempts in generating and observing coherent phonons, rigorous techniques to characterize and detect these phonon coherence in a crystalline material have been lagging compared to what has been achieved for photons. One main challenge is a lack of detailed understanding of how detection signals for phonons can be related to coherence. The quantum theory of photoelectric detection has greatly advanced the ability to characterize photon coherence in the last century and a similar theory for phonon detection is necessary. Here, we re-examine the optical sideband fluorescence technique that has been used detect high frequency phonons in materials with optically active defects. We apply the quantum theory of photodetection to the sideband technique and propose signatures in sideband photon-counting statistics and second-order correlation measurement of sideband signals that indicates the degree of phonon coherence. Our theory can be implemented in recently performed experiments to bridge the gap of determining phonon coherence to be on par with that of photons.
\end{abstract}

\maketitle

Phonons are packets of vibrational energy that shares similarity with its bosonic cousin: photons. Advances in nanofabrication has enabled many parallel developments in photon and phonon control. Parallel developments in passive control techniques include photonic \cite{Joannopoulos1997} versus phonoic crystals \cite{III2009}, optical \cite{Cai2010} versus acoustic metamaterials \cite{Ma2016} etc.  Development in active manipulation of electromagnetic waves have led to creation of nanoscale optical emitters \cite{Willander2014} and gates \cite{Chen2013} and similar progress have been made in controlling phonons especially in the realms of optomechanics \cite{Aspelmeyer2014} and phononic devices \cite{Li2012,Han2015}. Phonons span a vast frequency range and while techniques to control and sense lower frequency coherent phonons have been well-developed \cite{Ikezawa2001,Lanzillotti-Kimura2007,Vahala2009,Grimsley2011,Hong2012,Tian2014,Wang2014,Yoshino2015,Volz2016,Shinokita2016}, heat carrying coherent terahertz acoustic phonons have been harder to detect due to their small wavelength and numerous scattering mechanism \cite{Chen2005}. Recently, ultrafast surface deflection techniques have been used to generate and detect THz phonons \cite{Kent2002,Kent2006,Cuffe2013,Maznev2013}. At the same time, nanoscale material structures have exhibited phonon coherence through their thermal conductivity \cite{Luckyanova2012,Ravichandran2014,Latour2014,wang_decomposition_2014,Alaie2015,mu_ultra-low_2015,mu_coherent_2015}. Earlier, THz crystal phonons have been generated and detected in low temperature experiments using dopants \cite{Renk1971,Renk1979,Bron1980} or sideband detection \cite{mostoller_phonon_1971,Bron1977,Renk1979,Bron1980,schwartz_generation_1985,wybourne_phonon_1988}. Sideband detection is attractive compared to both thermal conductivity measurements and optical deflection techniques due to its ability to directly access atomic length scales where THz phonon wavelength couple with lattice phonons. Furthermore, sideband signals are universal in almost all material with optically active defects with phonon-induced broadening. Last but not least, the energy of the phonons detected can be precisely resolved using a optical spectrometer, allowing for a precise and yet broadband phonon detection. 

In light of the success using sideband detection techniques for high frequency phonon detection, we examine the potential for this technique to measure the coherence properties of these phonons. Sideband spectroscopy has been widely used as a means to detect non-equilibrium phonon population \cite{schwartz_generation_1985}, phonon propagation \cite{schwartz_generation_1985}, phonon transmission through interfaces\cite{Bron1977}, phonon band structure \cite{mostoller_phonon_1971,Alkauskas2014} etc. Extensive theoretical studies in the nineteen fifties have allowed rigorous understanding of sideband lineshape and have been utilized to complement neutron scattering phonon band structure measurements \cite{kuhner_lattice_1972,man_degree_1976}. Here, we use the basis of quantum theory of photodetection to rigorously characterize the phonon coherence using sideband detection. We conclude that sideband detection does not give the same detected quantity of intensity in photodetection, but a non-trivial function that depends on the phonon intensity. Using this function, we determine that sideband counting statistics can distinguish between a coherent phonon and a thermal phonon distribution. Then, we propose an experimental scheme similar to the Hanbury Brown Twist (HBT) experiment \cite{brown_correlation_1956} to obtain a second order correlation function of the sideband signal. This function shares similar features with the familiar intensity correlation function in HBT and can be used to determine whether the detected phonons are coherent. 

\begin{figure}[htbp]
\includegraphics[width=\textwidth]{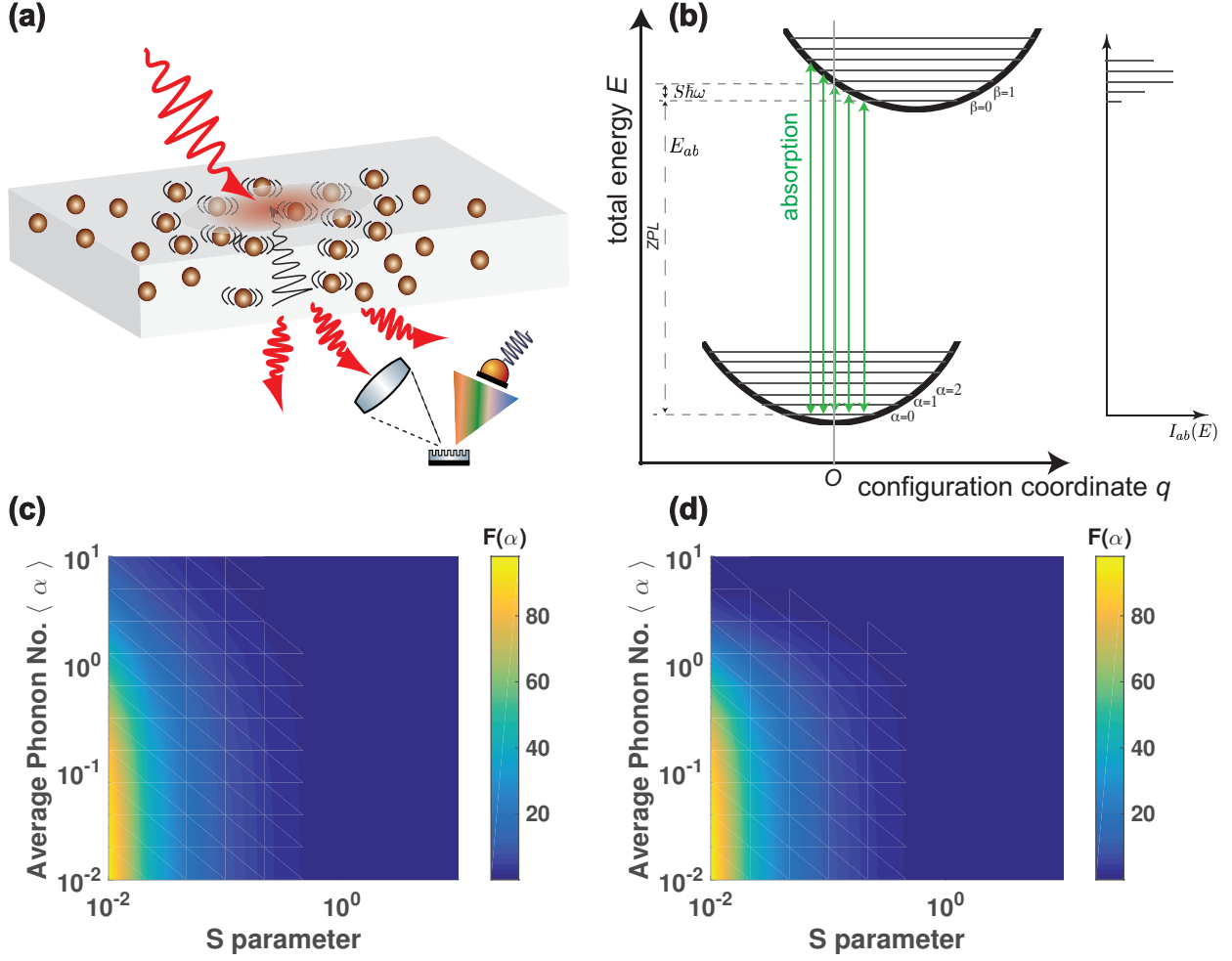}
\caption{\label{fig:1} (a) Schematic of sideband detection scheme. An incident optical beam gets absorbed and interacts with the local phonon population. The measured spectral properties of the transmitted photons with a grating spectrometer can give information of local phonon population. (b) Energy level diagram of a defect electron with vibrational transitions for the ground state and excite state. Absorption of a photon couples the ground to the excited state where the zero-phonon line (ZPL) couples the same vibrational levels ($\alpha=\beta$) while any other case $\alpha\neq\beta$ implies that phonons are absorbed. The S parameter determines the average phonon energy absorbed. The configurational coordinate $q$ represents displacement of a vibrational mode as a result of the normal mode transformation in Eq. S4. The spectrum on the right is a example of an absorption spectrum described by Eq. S11. (c) $\ave{F(\alpha)}$ versus S parameter and average phonon number $\ave{\alpha}$ for a thermal state ensemble. (d) $\ave{F(\alpha)}$ versus S parameter and average phonon number $\ave{\alpha}$ for a coherent state ensemble. }
\end{figure}

Figure \ref{fig:1}(a) shows the schematic of the sideband detection setup consisting of a crystal with defects. An incident optical beam excite the electrons in the defects, which absorbs some of the photons. The transmitted light gets collected into a grating spectrometer and then diffracted for frequency dependent detection. Figure \ref{fig:1}(b) shows the energy level diagram for an optically active defect \cite{Toyozawa2003}. The ground state defect electron resides in a harmonic potential consisting of different vibrational levels. The vibrational level the electron occupies is dictated by its equilibrium temperature. The excited state is assumed to have the same vibrational frequency. Under conditions of zero strain, electrons of the defects don't couple to the phonons in the crystal and the ground and excited state potentials will have a same minimum position ($q=0$ in Fig. \ref{fig:1}(b)). Optical excitation will thus only couple states of the same vibrational energy $\alpha=\beta$ between the ground and excited state and this is know as the zero-phonon line (ZPL) \cite{keil_shapes_1965}. However, defects introduce local strains which causes defect electrons to couple to phonons in the crystal through electron-phonon coupling \cite{lax_franckcondon_1952}. This leads to a shift in the excited state minimum which caused sidebands to appear about the ZPL. The additional energy between the ground state minimum to the excited state assuming a Franck-Condon (FC) optical transition is determined by the Huang-Rhys parameter or the S parameter. Relevant information on sideband theory from Refs. \cite{lax_franckcondon_1952,keil_shapes_1965,Toyozawa2003} are summarized in Section I of the Supplementary Information (SI). Here, we proceed with an important result derived from Eqs. S6-S9 of the sideband theory which describes the overlap strength between the ground vibrational state $\ket{\alpha}$ to the excited state $\ket{\beta}$ as

\begin{equation}
\lvert \Theta_{\alpha\beta} \rvert^2=\exp(-S_i)\frac{\alpha !}{\beta !} S_i^{\beta-\alpha} \lvert L_{\alpha}^{\beta-\alpha}(S_i) \rvert^2 \label{eq:overlap}
\end{equation}
where $S_i$ is the S parameter for the $i$th normal mode and the localized excitation and $L_{n}^{m}(z)$ are Laguerre polynomials \cite{keil_shapes_1965}. Equation \ref{eq:overlap} is used to explain shapes of absorption or emission spectra in sideband detection experiments \cite{mostoller_phonon_1971}. 

 At low temperature, almost all absorption processes will be stokes i.e. the absorbed photon be of higher energy than the ZPL \cite{wybourne_phonon_1988}. However, phonons do not equilibrate so easily at low temperature due to a lack of temperature-dependent scattering processes. Thus, non-equilibrium (NE) phonons generated can be detected at millimeters away from the source \cite{schwartz_generation_1985}. Nevertheless, no rigorous coherence characterizations has been carried out so far except from broadening inferences \cite{schwartz_generation_1985}. Here, we introduce the quantum theory of photoelectric detection (recapped in Section II of the SI) to evaluate the sideband signal for characterizing NE phonon coherence. To proceed, we would like to highlight an important result for the probability of one-photon photodetection at time $t_0+\Delta t$ 

\begin{equation}
P(t_0+\Delta t)\sim\eta \langle \ald\al\rangle \Delta t \label{eq:prob_photon}
\end{equation}
where $\ave{}$ means ensemble average, $\al,\ald$ are photon annihilation and creation operators such that $\ave{\ald\al}=I$ and $\Delta t$ is the measurement duration. Equation \ref{eq:prob_photon} tells us that the probability of detecting a photon is directly proportional to its intensity. 

    We apply the same formalism to study sideband detection of phonons. Physically, two processes happen in sideband detection compared to a single process of electron photon interaction. Firstly, there exist a electron-phonon between the ground state electron in the defect and the crystal such that any change in local phonon population is reflected in the defect phonon state \cite{henderson_optical_1989,Toyozawa2003}. Second, the ground state electron gets excited by electromagnetic wave to the excited state in which an absorption process happens \cite{henderson_optical_1989,Toyozawa2003}. Thus, an instantaneous sampling of phonon state happens when optical excitation occurs. In this work, we will focus on the one-phonon anti-stokes absorption sideband which has been used for phonon detection \cite{Bron1977}. The one-phonon ($\beta=\alpha-1$) overlap strength can be obtained from Eq. \ref{eq:overlap} as $F(\alpha)=\sum_{\beta}\lvert\bra{\beta} \bj \ket{\alpha}\rvert^2=\exp(-S_i)\alpha \lvert L_{\alpha}^{-1}(S_i) \rvert^2 S_i^{-1}$, where $\alpha$ is the ground vibrational state energy level in Fig. \ref{fig:1}(b). Figures \ref{fig:1}(c) and (d) show the ensemble averaged $\ave{F(\alpha)}$ a function of the average phonon number and S parameter for a coherent and thermal distribution. Note that $F(\alpha)$ is not a linear function with respect to the phonon intensity due to the oscillatory behavior of the Laguerre polynomials in Eq. \ref{eq:overlap}. While the differences between Figs. \ref{fig:1}(c) and (d) are subtle, they do lead to observable differences in the detected sideband signal as will be demonstrated shortly.
    
    Phonon fields can be expressed in a quantized form like the photon fields \cite{Toyozawa2003,wagner_investigation_2005} which describes electron phonon coupling (see Eq. S20 of SI). We formulate the interaction Hamiltonian for the crystal phonon to defect electron's phonon interaction as
    \begin{equation}
    H_I=\hat{d}\xi_l \hat{A}_l\sum_j\Lambda_j \bt{} \label{eq:ham_phonon}
    \end{equation}
     
where $\hat{d}$ is the dipole operator,  $\hat{A}_l=\al e^{-i\olm t_1} +\ald e^{i\olm t_1} $ such that $\hat{a}$ is the photon annihilation operators for photon absorption with frequency $\omega_l$, $\hat{B_j}=(\hat{b}_j\exp(i (k_j r_0-\omega_j t))+\hat{b}^{\dagger}_j \exp(-i (k_j r_0-\omega_k t)))$ for defect phonon operators of the $j$th modes \cite{mostoller_phonon_1971}, and $\xi_l$ and $\Lambda_j$ are the product for the proportionality constant of the photon and phonon field (defined in SI), respectively. This derivation is one-dimensional but can be easily generalized to three dimension. Also, we do not consider the polarization dependence of the lattice modes or the photon fields in this derivation. 
    
    We now relate to Eq. \ref{eq:overlap} for absorption spectrum to the anti-stokes one-phonon ovelap $F(\alpha)$ . After some algebra (Eq. S33 of SI), our sideband detection probability in Eq. S28 is simplified to 
    
    \begin{equation}
    P(t,\Delta t)=\eta F(\alpha) \Delta t \delta(\olm-(\ob-\oa)+\oj) \label{eq:prob_phonon_simp}
    \end{equation}

    \begin{figure*}
    \centering
     \includegraphics[width=\textwidth]{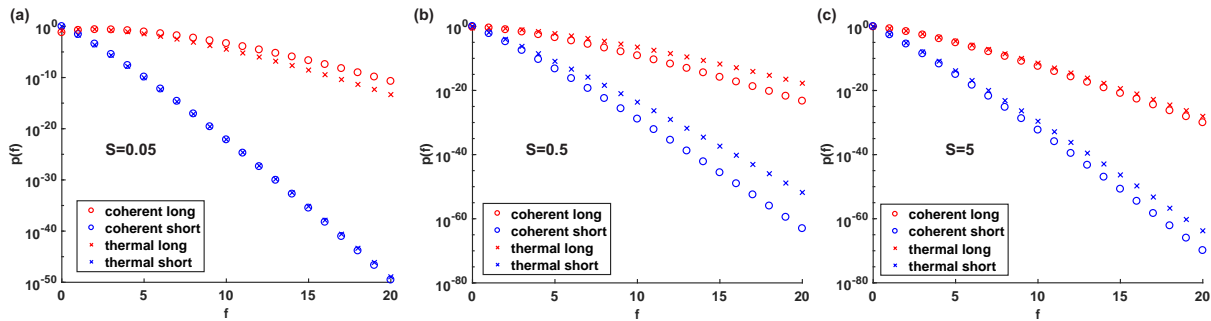}
    \caption{\label{fig:2} (a-c) Plots of Eqs. \ref{eq:n_phonon_long} and \ref{eq:n_phonon_short} labeled as ``long" (red)  and``short" (blue) respectively for different S parameters. Coherent state ensemble is labeled with circle and thermal state ensemble is labeled with cross. }
    \end{figure*}
    
    We can use Equation \ref{eq:prob_phonon_simp} to obtain the $n$ particle detection probability. In the optical case, $n$ photon detection probability is derived from one-photon detection probability in Eq. \ref{eq:prob_photon} such that \cite{kelley_theory_1964}
        \footnotesize
    \begin{equation}
    p(n,t,T)=\frac{1}{n!}\left(\eta \int_t^{t+T} I(t') dt'\right)^n \exp\left( -\eta \int_t^{t+T} I(t') dt' \right) \label{eq:n_photon}
    \end{equation}
\normalsize
 where $T$ is the detection time and $\eta$ is the overall detection efficiency.  Equation \ref{eq:n_photon} shows that the probability of detection follows a Poisson distribution with the expected photon number $\ave{n}=\eta \int_t^{t+T} I(t') dt'$. In the long-time limit,  $\ave{n}\approx\eta  \ave{I} T $ and we obtain \cite{mandel_optical_1995}
    
    \begin{equation}
    p(n,t,T)\approx\frac{1}{n!}\left(\eta \ave{I} T \right)^n \exp\left( -\eta \ave{I} T \right) \label{eq:n_photon_long}
    \end{equation}
    
    Whereas in the short-time limit, $\ave{n}\approx\eta  I(t) T $ and \cite{mandel_optical_1995}
    
    \begin{equation}
    p(n,t,T)=\int_0^{\infty}\frac{1}{n!}\left(\eta I T \right)^n \exp\left( -\eta I T \right) P(I) dI \label{eq:n_photon_short}
    \end{equation}
     where the photon intensity $I$ now is a random variable. The specific form of $P(I)$ depends on the distribution of the photon ensemble.  In the coherent ensemble, Eqs. \ref{eq:n_photon_long} and \ref{eq:n_photon_short} be the same but not for a thermal ensemble \cite{mandel_optical_1995}. 
          
    Now, let us consider the case of phonon sideband detection where we are not directly measuring the phonon intensity but a function of the intensity described by Eq. \ref{eq:prob_phonon_simp}. Let us define $f=F(\alpha)$ so that the averaged measurable quantity $\ave{f}=\eta \int_t^{t+T} F(\alpha(t')) dt'$, where the ground vibrational state $\alpha$ is the time-dependent. Then, the f-probability can be formulated as a Poisson distribution just like Eq. \ref{eq:n_photon} to be 
    \footnotesize
    \begin{equation}
    p(f,t,T)=\frac{1}{f!}\left(\eta \int_t^{t+T} F(\alpha(t')) dt'\right)^f \exp\left( -\eta \int_t^{t+T} F(\alpha(t')) dt' \right) \label{eq:n_phonon}
    \end{equation}
    \normalsize
    Let us take a look at the short and long time limit of $p$. For long time limit, $\ave{f}\approx\eta  \ave{F} T $ such that 
    
    \begin{equation}
    p(f,t,T)\approx\frac{1}{f!}\left(\eta \ave{F} T \right)^f \exp\left( -\eta \ave{F} T \right) \label{eq:n_phonon_long}
    \end{equation}
    
    just like Eq. \ref{eq:n_photon_long} except that it is not a direct function of intensity. In the short time limit, $\ave{f}_{\alpha}\approx\eta  F(t) T $ but  
    
    since $P(\alpha)$ is a discrete distribution of the phonon ensemble, we sum over all states $\alpha$ to obtain
     
     \begin{align}
    p(f,t,T)=&\sum_{\alpha}\frac{1}{f!}\left(\eta F(\alpha) T \right)^f \exp\left( -\eta F(\alpha) T \right) P(\alpha) \\
    =&\ave{\frac{1}{f!}\left(\eta F T \right)^f \exp\left( -\eta F T \right)}_{\alpha} \label{eq:n_phonon_short}
    \end{align}
     
    The difference between Eq. \ref{eq:n_phonon_short} and Eq. \ref{eq:n_photon_short} is in the measured variable but they share the same physical principle where short-time limit fluctuations arises due to ensemble average of the whole detection probability rather than just the measured quantity. Here, we note the similar feature where a coherent state will lead to the same result in the short and long time limit (Eqs. \ref{eq:n_phonon_short} and \ref{eq:n_phonon_long}) but not for thermal state.
    
    Figure \ref{fig:2} shows the probability distribution $p(f)$ for different time scales $\eta T$ under a coherent or thermal ensemble. Here we choose different values of $\eta T$ to account for the short ($\eta T=0.1$) to the long time limit ($\eta T=10$). Under these cases, we obtain two limits for the coherent ensemble and compare the thermal ensemble for different values of S parameter. First, we notice that the S parameter strongly affects the sensitivity of the difference between a coherent and a thermal state. A small S parameter will lead to virtually undetectable difference between the two. Physically, a small S parameter means less average phonons absorbed or emitted and thus less sensitivity. Second, higer values of $S$ does not imply greater difference in signal between coherent and thermal states. This is caused by the oscillatory nature of the Laguerre polynomials in Eq. \ref{eq:overlap}. Third, coherent and thermal states show different probability in both the short and long time limit, unlike the case (in Eq. \ref{eq:n_photon_long}) for photon intensity. This is because the function $\ave{F(\alpha)}$ will have different average values in Eq. \ref{eq:n_phonon_long} (evident from Figs. \ref{fig:1}(c,d)) between a thermal and a coherent state for the same average phonon number even in the long time limit. Photon probability measurements have been used to characterize coherent and non-classical states \cite{arecchi_measurement_1965} and we can likewise carry out similar measurements for sideband phonon detection.
 
 \begin{figure}[htbp]
    \centering
    \includegraphics[width=\textwidth]{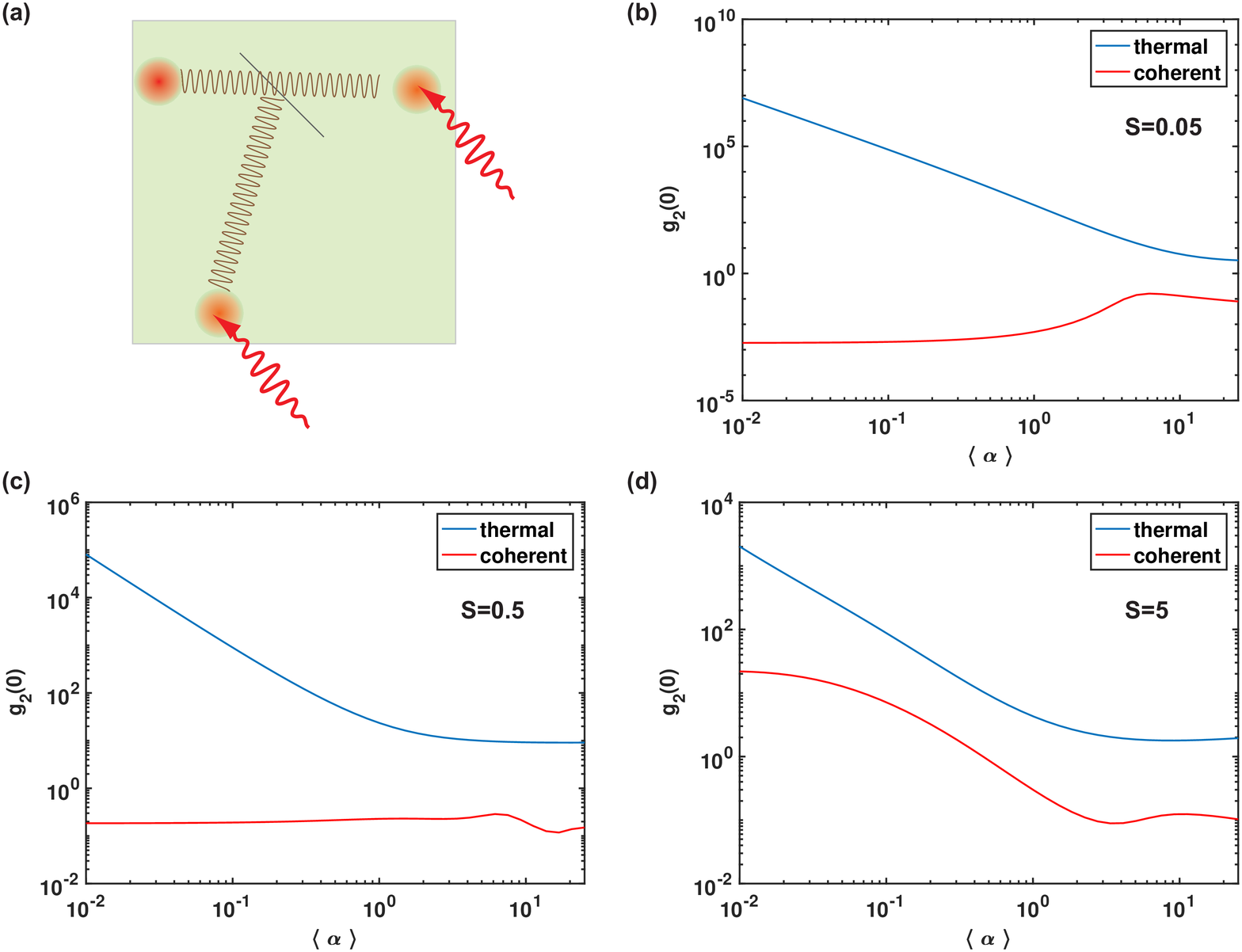}
    \caption{\label{fig:3} (a) Schematic of proposed measurement of second order phonon correlation using sideband detection. An optical excitation generates a phonon beam which gets split by a clean interface into two paths. Detection on each path with sideband technique allows us to measure the second order correlation defined in Eq. \ref{eq:g2_phonon}. (b-d) Second order sideband correlation for different values of S parameter and average phonon number $\langle \alpha \rangle$ at $\tau=0$. The red line indicates thermal ensemble and blue indicates coherent ensemble. Notice that thermal state is always above the coherent state for all values of average phonon number $\ave{\alpha}$ and S parameters.}
    \end{figure}
    A more common technique to characterize optical coherence is using a HBT experiment \cite{brown_correlation_1956,mandel_optical_1995}. We propose a similar setup in Fig. \ref{fig:3}(a) for sideband phonon detection. Here, an optical excitation generates a phonon beam which becomes incident on a smooth oblique interface and gets partially reflected and transmitted. Two optical beams incident on different locations of the crystal are then used to create two sideband signals which are correlated with each other. Before examining this proposal further, let us first recap how the photon intensity correlation function looks like
    
    \begin{equation}
    \second(t,t+\tau)=\frac{\ave{\ald(t)\ald(t+\tau)\al(t+\tau)\al(t)}}{\ave{\ald(t)\al(t)}^2} \label{eq:g2_photon}
    \end{equation}
    
    The numerator in the intensity correlation in Eq. \ref{eq:g2_photon} is ordered with $\ald$ before $\al$ and time-ordered such that a state $\bra{\alpha}\ald\ald\rightarrow\bra{\alpha-2}$ (same for $\al\al\ket{\alpha}$), which implies two consecutive absorption processes by two photodetectors from the same state. The denominator just implies independent absorption events. For our case, we are measuring $F(\alpha)$ which projects the phonon intensity onto a function and prevents us from defining our second order correlation in the same manner as in Eq, \ref{eq:g2_photon}. Nevertheless, we can define a second order correlation that describes the same physical process of correlated absorption events versus independent events. In this case, we have 
    \begin{widetext}
    \begin{align}
    \second(t,t+\tau)=&\frac{\sum_{\alpha}P(\alpha)\sum_{\beta,\beta'} \bra{\alpha} \bjd(t) \ket{\beta}\bra{\beta} \bj(t) \ket{\alpha} \bra{\alpha-1} \bj(t+\tau) \ket{\beta'} \bra{\beta'} \bjd(t+\tau) \ket{\alpha-1}   }{\ave{F(\alpha(t))}^2_{\alpha}}\\
    =&\frac{\ave{ F(\alpha(t))F(\alpha(t+\tau)-1)}_{\alpha}}{\ave{F(\alpha(t))}^2_{\alpha}} \label{eq:g2_phonon}
    \end{align}
    \end{widetext}

    Figures \ref{fig:3} plot the second order correlation for sideband detection defined in Eq. \ref{eq:g2_phonon} at $\tau=0$ for different values of S parameters. Notice that we have the thermal ensemble being above the coherent ensemble for all values of average phonon intensity $\ave{\alpha}$ for different values of S parameter. This is in agreement with the behavior of photon intensity correlation at time $\tau=0$ in Eq. \ref{eq:g2_photon}. However, photon intensity correlation does not vary with the average photon intensity or other parameters unlike our sideband correlation here, which can vary by orders of magnitude for different S parameter and average phonon numbers. 
    
Our work differs from those in optomechanics and non-linear coherent phonon control \cite{kozak_coherent_2013}. Optomechanics primarily relies on coupling a mechanical mode to a designed optical cavity for coherent phonon control. It is remarkable that quantum coherence of phonons has been predicted \cite{hu_quantum_1996,hu_phonon_1999,hu_quantum_2015} and observed \cite{Aspelmeyer2014} in this field. Here, we are proposing sideband detection with optical defects which couples to intrinsic phonon modes in materials. Also, we only restrict our discussion here to coherent and thermal state although it is possible to consider other quantum states such as Fock states and squeezed states \cite{hu_quantum_1996,hu_phonon_1999,hu_quantum_2015} . For non-linear coherent phonon generation, an optical field directly couples to optical phonons \cite{kozak_coherent_2013} or zone-center acoustic phonons \cite{olsson_temperature_2015} and phase matching always results in coherent phonons. Our work actually detects high frequency acoustic phonons which are not capable of direct coupling to light through phase matching. Furthermore, our technique can detect both coherent and incoherent phonons through their ensemble distribution and no phase matching is required. Quantum mechanics has been used to describe THz phonons such as in nitrogen vacancy phonon mediated gates \cite{Albrecht2013}, controlling quantum phonon states in materials \cite{hu_quantum_1996,hu_phonon_1999,hu_quantum_2015} and phonon coherence in thermal transport \cite{Latour2014} but not in the same way as a detection scheme illustrated here.

    The feasibility of our proposals are certainly within current experimental reach. Photon counting statistics is a very established field and can be easily applied to sideband detection experiments. Measuring phonon sideband correlation is achievable given the low temperature of the experiment and ability to shape and make good quality crystal interfaces \cite{ichinose_crystal_2000}. Last but not least, small nano crystals can be used as local phonon detectors similar to Ref. \cite{Tian2014,laraoui_imaging_2015} where sideband signals can be used to determine phonon coherences in optically-inert materials of interest.
    
    In conclusion, we have applied the quantum theory of photodetection to the sideband technique and propose signatures in sideband photon-counting statistics and second-order correlation of the sideband signal that phonons are in a coherent state. Our theory can be implemented experimentally to bring the ability of characterizing phonon coherence to be on par with that of photons.

\begin{acknowledgments}
\end{acknowledgments}
%

\end{document}